\documentclass[prd,twocolumn,showpacs,preprintnumbers,aps,nofootinbib]{revtex4-2}

\usepackage{graphicx}  
\usepackage{dcolumn}
\usepackage{bm}
\usepackage{amsmath,amssymb,amsfonts}
\usepackage{latexsym}
\usepackage{color}
\usepackage[colorlinks=true,urlcolor=black,anchorcolor=blue
,citecolor=blue,filecolor=blue,linkcolor=blue,menucolor=blue
,linktocpage=true,pdfproducer=medialab,pdfa=true]{hyperref}
\def\nn    {\nonumber}

\begin{document}
\title{\boldmath
Probing the general 2HDM with flavor violation through $A \to ZH$}
\author{Wei-Shu Hou and Mohamed Krab}
\affiliation{Department of Physics, National Taiwan University, Taipei 10617, Taiwan}
\bigskip
\begin{abstract}
We investigate the LHC discovery prospects for a second Higgs doublet through $A \to ZH$ weak decay. The latter is identified as the {\it smoking gun signature} of two Higgs doublet models (2HDMs) with first-order electroweak (EW) phase transition, a necessary condition for EW baryogenesis.
In the general 2HDM (G2HDM) that has flavor-changing neutral Higgs couplings, $H$ may decay dominantly via $t\bar c + \bar tc$ final states, giving rise to trilepton signals. By a phenomenological analysis, we show that $A \to ZH$ in $\ell^+ \ell^-t\bar c$ or $\ell^+ \ell^-\bar t c$ final states could be a promising probe of G2HDM at the LHC with flavor violation. 
\end{abstract}

\maketitle
%

\section{Introduction}
The existence of a second Higgs doublet introduces extra Higgs bosons beyond the one discovered at the LHC~\cite{ATLAS:2012yve,CMS:2012qbp}, and could provide the mechanism to explain the matter-antimatter asymmetry of the Universe. Its discovery would confirm the nonminimal structure of the Higgs sector and perhaps shed light on the mechanism behind the baryon asymmetry of the Universe (BAU).
	
We specifically advocate the general two Higgs doublet model (G2HDM), i.e. without the usual $Z_2$ symmetry. 
With two identical weak doublets, it allows for a second set of Yukawa couplings, other than fermion masses. Though it has not garnered much attention, this model has quite a few merits. 

Quartic couplings at ${\mathcal O}(1)$ can provide~\cite{Kanemura:2004ch} prerequisite first-order electroweak phase transition (FOEWPT), needed for electroweak baryogenesis (EWBG), while ${\mathcal O}(1)$ extra top Yukawa couplings $\rho_{tc}$ and $\rho_{tt}$ can {\it each} drive~\cite{Fuyuto:2017ewj} EWBG to account for BAU, one of the biggest mysteries. 
(See also Refs.~\cite{Turok:1990zg,Davies:1994id,Cline:1995dg,Cline:1996mga,Fromme:2006cm,Cline:2011mm,Ahmadvand:2013sna,Dorsch:2013wja,Dorsch:2014qja,Basler:2016obg,Bernon:2017jgv,Chiang:2016vgf,Dorsch:2016nrg,Dorsch:2017nza,Modak:2018csw,Goncalves:2021egx,Basler:2021kgq,Enomoto:2021dkl,Kanemura:2023juv,Goncalves:2023svb,Athron:2025iew,Biekotter:2023eil} for FOEWPT and EWBG in the 2HDM.)
Given this, one must confront the stringent electron electric dipole moment bounds of ACME~\cite{ACME:2018yjb} and JILA~\cite{Roussy:2022cmp}, where a spectacular cancellation mechanism~\cite{Fuyuto:2019svr} allows one to survive these bounds. 

Without the $Z_2$ symmetry, G2HDM allows for flavor-changing neutral
Higgs (FCNH) couplings, such as $t \to ch$~\cite{Hou:1991un}. Interestingly, this decay still has not been seen, as it is suppressed by {\it alignment}, by a rather small $h$-$H$ mixing between the two {\it CP}-even scalars. However, FCNH couplings can lead to {\it alignment enhanced} processes such as $cg \to tH/tA \to tt\bar c, tt\bar t$~\cite{Kohda:2017fkn}, while processes such as $cg \to bH^+$~\cite{Ghosh:2019exx} and $\bar b g \to \bar cH^+$~\cite{Hou:2024bzh}, along with $c\bar b \to H^+$~\cite{Hou:2024ibt} resonant production, are not subject to Cabibbo-Kobayashi-Maskawa (CKM) suppression as in the 2HDM with $Z_2$ symmetry.

In this paper, we study the discovery prospects of $A \to ZH$ decay in  $\ell^+\ell^- t\bar c$ or $\ell^+\ell^- \bar t c$ (henceforth, $\ell^+\ell^- tc$) final states. This decay channel is identified as the smoking gun signature at the LHC of 2HDM scenarios with FOEWPT~\cite{Dorsch:2014qja,Basler:2016obg,Bernon:2017jgv,Biekotter:2023eil}.
We propose a benchmark scenario and perform a detector level Monte Carlo analysis at the 14~TeV LHC. We show that searches for the $A \to ZH$ decay into the $\ell^+\ell^- t c$ final state are promising: Run 2 data may already lead to discovery, thus providing a probe of G2HDM that has flavor-violating couplings.

\section{G2HDM}
In the Higgs basis where only one doublet gives mass, the most general {\it CP}-conserving scalar potential for two doublets $\Phi$ and $\Phi'$ is given by~\cite{Davidson:2005cw,Hou:2017hiw}
\begin{align}
 & V(\Phi,\Phi') = \mu_{11}^2|\Phi|^2 + \mu_{22}^2|\Phi'|^2 \nn
    - (\mu_{12}^2\Phi^\dagger\Phi' + \rm{H.c.}) \label{pot}\\
 & \quad + \frac{\eta_1}{2}|\Phi|^4 + \frac{\eta_2}{2}|\Phi'|^4
   + \eta_3|\Phi|^2|\Phi'|^2  + \eta_4 |\Phi^\dagger\Phi'|^2 \nn \\
 & ~ + \left[\frac{\eta_5}{2}(\Phi^\dagger\Phi')^2
   + \left(\eta_6 |\Phi|^2 + \eta_7|\Phi'|^2\right) \Phi^\dagger\Phi' 
   + \rm{H.c.}\right],
\end{align}
where all mass and quartic coupling parameters are chosen to be {\it real}, $\Phi$ generates the vacuum expectation value $\mathit{v}$ for electroweak (EW) symmetry breaking through a first minimization condition of $\mu_{11}^2 = - \frac{1}{2}\eta_1 v^2$, while $\left\langle \Phi'\right\rangle = 0$; hence $\mu_{22}^2 > 0$. A second minimization condition $\mu_{12}^2 = \frac{1}{2}\eta_6 v^2$ reduces the number of parameters to 9.

In the Higgs basis, the {\it CP}-odd scalar $A$ and the charged Higgs $H^\pm$ are mass eigenstates. The {\it CP}-even mass eigenstates $h$ and $H$ are obtained by diagonalizing the mass-squared matrix through a rotation characterized by the angle $\gamma$, denoted as $\beta-\alpha$ in 2HDM with $Z_2$ symmetry convention, which satisfies $s_\gamma c_\gamma = \eta_6 v^2/(m^2_H-m^2_h)$~\cite{Davidson:2005cw,Hou:2017hiw}, with $c_\gamma \equiv \cos\gamma$ ($s_\gamma \equiv \sin\gamma$).  
The emergent alignment phenomenon, that $h$ closely resembles the SM Higgs boson~\cite{ATLAS:2016neq}, implies a rather small $h$-$H$ mixing angle $c_\gamma$.

The general Yukawa couplings are \cite{Davidson:2005cw,Hou:2017hiw}
\begin{align}
 \mathcal{L}_Y = 
 & - \frac{1}{\sqrt{2}} \sum_{f = u, d, \ell}
       \bar f_{i} \bigg[\big(-\lambda^f_{ij} \sin\gamma + \rho^f_{ij} \cos\gamma\big)h \nn \\
 & + \big(\lambda^f_{ij} \cos\gamma + \rho^f_{ij} \sin\gamma\big) H
   -i\,{\rm sgn}(Q_f) \rho^f_{ij} A \bigg]  R f_{j}\nn \\
 & - \bar{u}_i\big[(V\rho^d)_{ij} R - (\rho^{u\dagger}V)_{ij} L\big]d_j H^+ \nn \\
 & - \bar{\nu}_i\rho^\ell_{ij} R \ell_j H^+ +{\rm H.c.},
\label{LYukawa}
\end{align}
where $i,j = 1, 2, 3$ are the generation indices, $\lambda^f_{ij} \equiv \delta_{ij}\sqrt{2}m_i^f/v$, ${\rm sgn}(Q_f) = +1(-1)$ for $f=u$ ($f=d,\ell$), $L,R = (1\mp\gamma_5)/2$, and $V$ is the CKM matrix. The $\lambda^f_{ij}$ matrices are real and diagonal, while $\rho^f_{ij}$ are in general complex and nondiagonal. Here, we consider a {\it CP}-conserving model and take the $\rho^f_{ij}$ matrices as real but not necessarily Hermitian.

In the alignment limit of $c_\gamma \to 0$, $h$ couples diagonally as the SM Higgs boson, while $H$ interacts through the extra Yukawa couplings $\rho^f_{ij}$. Thus, besides mass-mixing hierarchy protection~\cite{Hou:1991un} of FCNH couplings, alignment provides~\cite{Hou:2017hiw} further safeguard, such as for $t \to ch$, without relying on natural flavor conservation~\cite{Glashow:1976nt}. So, enforcing the $Z_2$ symmetry may be overkill. The $H^+$ and $A$ couplings are independent of $\gamma$.

\begin{figure}[t!]
	\centering
	\includegraphics[scale=0.45]{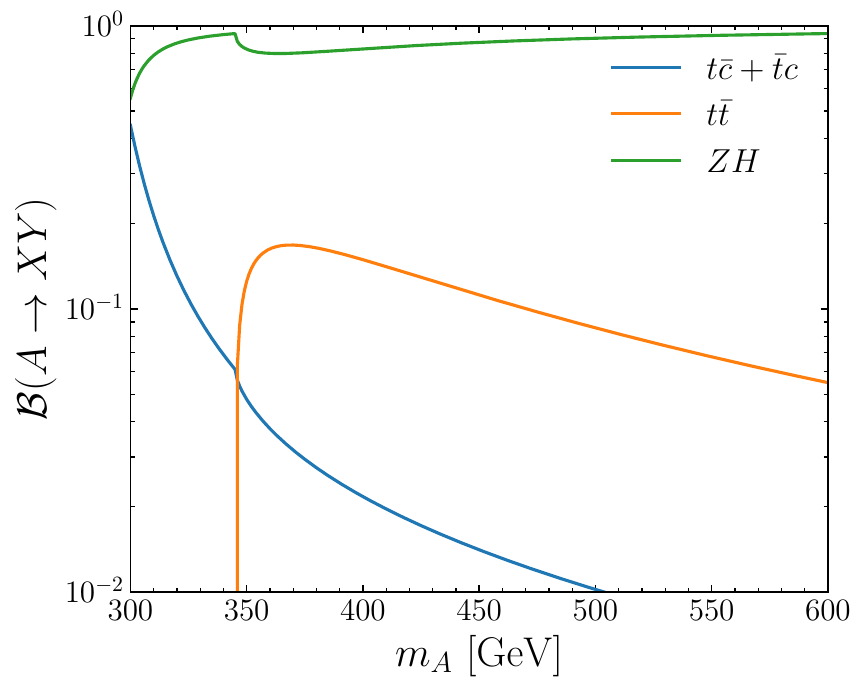}
	\caption{Branching ratios of the {\it CP}-odd scalar $A$ as a function of $m_A$ for $m_H = 200$~GeV, $\rho_{tt}=0.3$, and $\rho_{tc}=0.1$.}	 \label{fig:brAxy}
\end{figure}

In what follows, we drop the superscript $f$ and set all $\rho_{ij} = 0$ for simplicity, except $\rho_{tt}$ and $\rho_{tc}$. We will return to the effect of turning on other $\rho_{ij}$ later in the paper. We set $c_\gamma = 0$, and fix $m_h = 125$ GeV.
In Fig.~\ref{fig:brAxy}, we plot the branching ratios of $A \to t\bar c+ \bar t c,\,t\bar t,\,ZH$ for $m_H = 200$~GeV, $\rho_{tt}=0.3$, and $\rho_{tc}=0.1$. We assume $m_{H^+} \simeq m_A$ (custodial symmetry), which is required by EW precision observables, implying that the decay $A \to W^- H^+$ is kinematically forbidden. We will return to $m_{H^+} \simeq m_H$ (twisted custodial~\cite{Gerard:2007kn}), where $A \to W^- H^+$ could be of similar magnitude to $A \to ZH$ later. We observe that $A \to ZH$ largely dominates over $A \to t\bar c + \bar t c$ and $A \to t\bar t$. Note that $A \to t\bar c + \bar t c$ dominates $A$ decays for $m_A - m_H < m_Z$. The decay $A \to Zh$ is suppressed by $c_\gamma$. The subsequent decay of $H \to t\bar c+ \bar t c$ is dominant, while $H \to W^+W^-$ and $H \to ZZ$ decays are $c_\gamma$ suppressed. 
Above the $t\bar t$ threshold, $H \to t\bar t$ opens up and could be the dominant decay channel.
We note that a large $m_A$-$m_H$ splitting is still possible, such that FOEWPT is realized while the $t\bar t$ channel remains open; see, e.g., Refs.~\cite{Dorsch:2014qja, Basler:2016obg, Bernon:2017jgv, Biekotter:2023eil} and references therein.

%

\section{Constraints on G2HDM}
The model is subject to theoretical and experimental constraints. Higgs quartic couplings in Eq.~(\ref{pot}) are required to fulfill positivity, tree-level unitarity, and perturbativity constraints, which are implemented in \texttt{2HDMC-1.8.0}~\cite{Eriksson:2009ws}. 
G2DHM parameter space is further required to fulfill EW precision constraints through the oblique parameters $S$, $T$, and $U$~\cite{Grimus:2007if} using the Particle Data Group values~\cite{ParticleDataGroup:2024cfk}.
The code \texttt{2HDMC} calculates the theoretical predictions for these observables. We refer to Ref.~\cite{Hou:2024bzh} for details on the parameter space scan of the G2HDM and constraints.

\begin{figure}[t!]
	\centering
	\includegraphics[scale=0.45]{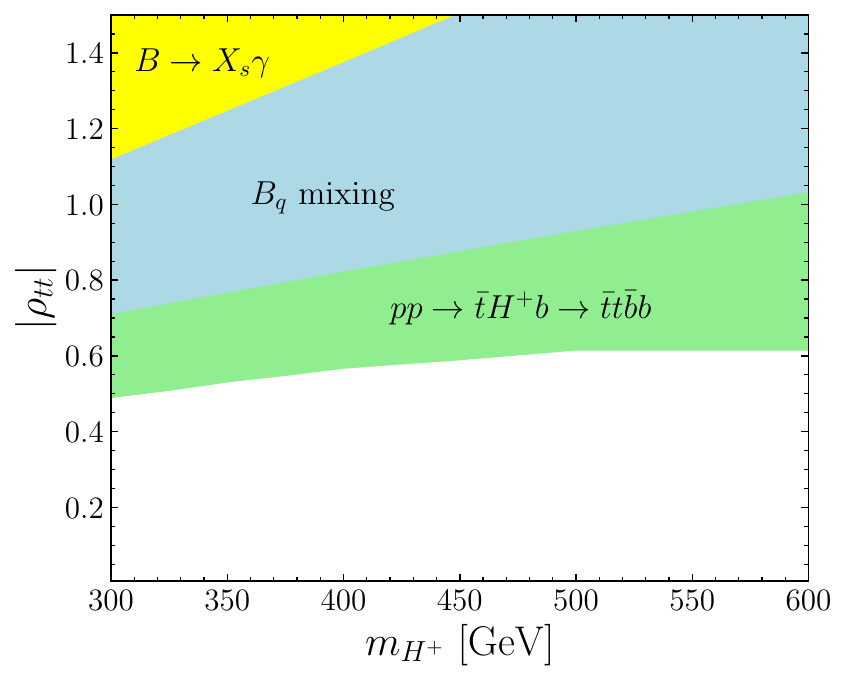}	
	\caption{Exclusion regions from $B \to X_s\gamma$, $B_q$ mixing, and ATLAS $pp \to \bar tH^+ b \to \bar t t\bar b b$ search in the $m_{H^+}$-$\left|\rho_{tt}\right|$ plane.}	\label{fig:rhott-bounds}
\end{figure}

Flavor constraints on extra top Yukawa couplings $\rho_{tt}$ and $\rho_{tc}$ are relatively mild~\cite{Crivellin:2013wna,Altunkaynak:2015twa}, where it is found that~\cite{Crivellin:2013wna,Altunkaynak:2015twa} $B_q$ mixing ($q = d, s$) strongly constrains $\rho_{ct}$ due to the CKM $|V_{cq}/V_{tq}| \sim 25$ enhancement factor; hence, $\rho_{ct}$ must be tiny. Therefore, $\rho_{ct}$ is turned off here.
Following Ref.~\cite{Altunkaynak:2015twa}, we estimate the modifications to the $B_q$-$\bar{B}_q$ mixing amplitude and use the UTfit results~\cite{UTFit2023}, $C_{B_d} = 1.09 \pm 0.09$ and $C_{B_s} = 1.10 \pm 0.06$, where $C_{B_q} \equiv M^q_{12}/M^q_{12}|^{\rm{SM}}$, while the phase $\phi_{B_q}$ is neglected in our study. We find the light blue region (extending to upper left) shown in Fig.~\ref{fig:rhott-bounds} excluded at $2\sigma$ level. The $B_q$ mixing constraint on $\rho_{tc}$ is weaker~\cite{Crivellin:2013wna}.
Reinterpreting the limits from Ref.~\cite{Crivellin:2013wna}, we find the bound $|\rho_{tc}| \lesssim 1.7$ for $m_{H^+} = 500$~GeV. 

The $B \to X_s\gamma$ constraint on $m_{H^+}$ in G2HDM is not particularly strong~\cite{Crivellin:2013wna,Altunkaynak:2015twa}.
Reference~\cite{Altunkaynak:2015twa} finds that $\rho_{bb}$ is more constrained than $\rho_{tt}$ due to an enhancement factor of $m_t/m_b$; hence, $\rho_{bb}$ must be small. Setting $\rho_{bb} = 0$, we find the yellow exclusion region depicted in Fig.~\ref{fig:rhott-bounds}. The $B \to X_s\gamma$ constraint on $\rho_{tc}$ is much weaker due to the small charm quark mass and the absence of any enhancement factor~\cite{Altunkaynak:2015twa}. 

LHC indirect and direct measurement set limits on $\rho_{tt}$, but these are weaker~\cite{Hou:2018uvr} due to $c_\gamma$ suppression [see Eq.~(\ref{LYukawa})]. We further check these constraints using the new version of the code \texttt{HiggsSignals}~\cite{Bechtle:2020uwn}, which incorporates the latest ATLAS and CMS Run 2 measurements, via \texttt{HiggsTools}~\cite{Bahl:2022igd}. 
ATLAS~\cite{ATLAS:2021upq} and CMS~\cite{CMS:2020imj} direct searches for $pp \to \bar t H^+ (b)$ followed by $H^+ \to t\bar b$ (charge conjugate process implied) constrain $\rho_{tt}$ as well.
These searches set significant limits on $\sigma[pp \to \bar t H^+ (b)]\times \mathcal{B}(H^+ \to t\bar b)$. We reinterpret ATLAS limits~\cite{ATLAS:2021upq} assuming $\mathcal{B}(H^+ \to t\bar b) = 100\%$ and obtain the light green exclusion region (extending upward) shown in Fig.~\ref{fig:rhott-bounds}. CMS limits~\cite{CMS:2020imj} are also reinterpreted and found to be weaker; thus, only the ATLAS bounds are shown.

\begin{figure*}[t!]
	\centering
	\includegraphics[scale=0.4]{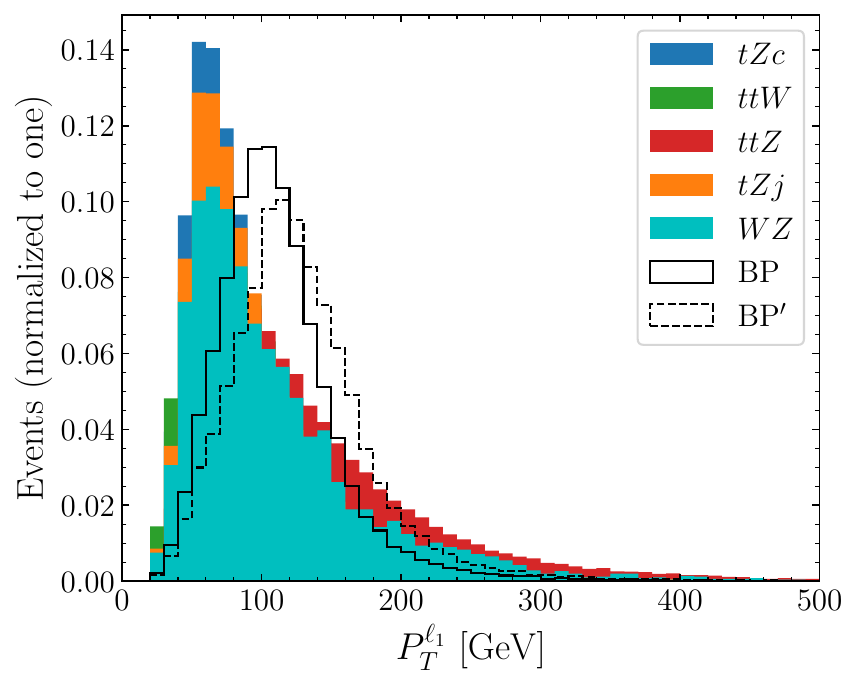}
	\includegraphics[scale=0.4]{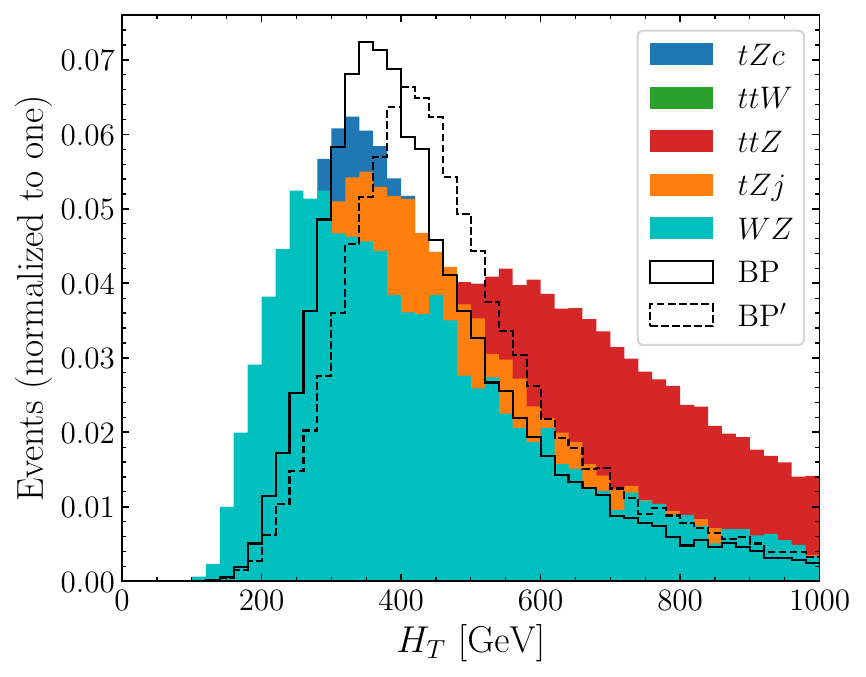}	
	\includegraphics[scale=0.4]{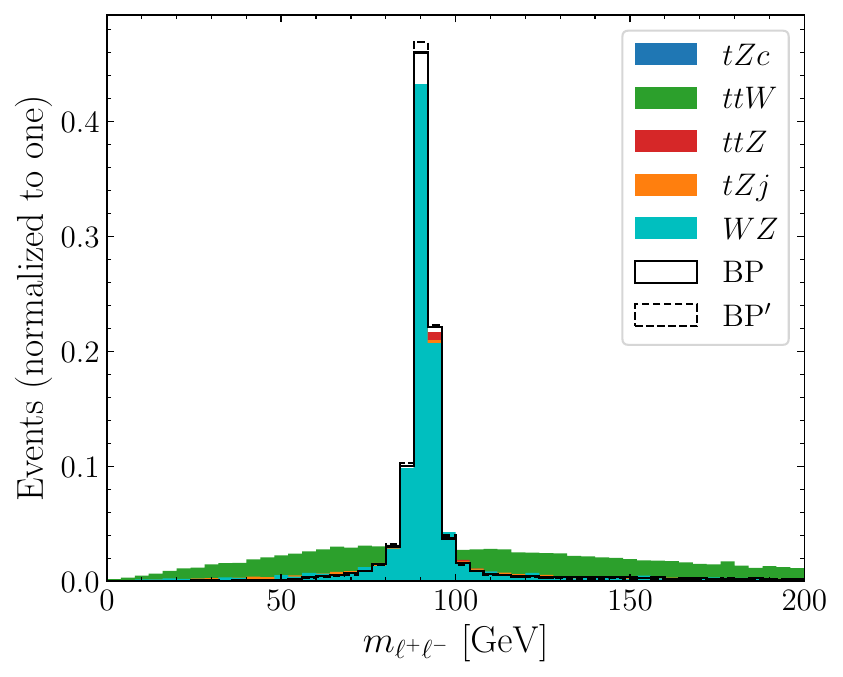}	
	\caption{$P^{\ell_1}_T$ (left), $H_T$ (middle), and $m_{\ell^+\ell^-}$ (right) distributions for background and signal processes. }	
	\label{fig:distr}
\end{figure*}

ATLAS~\cite{ATLAS:2023tlp} has performed searches for exotic Higgs bosons with flavor-violating couplings in final states with multiple leptons and $b$-tagged jets. 
The search results are interpreted in terms of a G2DHM with extra top Yukawa couplings $\rho_{tt}$, $\rho_{tc}$, and $\rho_{tu}$.
An additional scalar boson with masses $m_H$ between 200 and 620 GeV and couplings $\rho_{tt} = 0.4$, $\rho_{tc} = 0.2$, and $\rho_{tu} = 0.2$ is excluded at 95\% confidence level (CL).
CMS~\cite{CMS:2023xpx} searches for same-sign top associated production with a jet, via the $pp \to tH/A \to tt\bar c$ and $pp \to tH/A \to tt\bar u$ processes, set limits on $\rho_{tc}$ and $\rho_{tu}$.
In the absence of interference between $H$ and $A$, and setting $\rho_{tc} = 1.0$, $H$ or $A$ with masses below approximately 770 GeV is excluded at 95\% CL. Including interference effects extends the exclusion to $m_A = 810$~GeV at 95\% CL.
For our conservative choice of $\rho_{tc} = 0.1$, no limits are set.
For nonzero $c_\gamma$, LHC $t \to qh$ ($q = u, c$) searches~\cite{CMS:2021hug,ATLAS:2024mih} set further limits on $\rho_{tc}$ and $\rho_{tu}$.
For $c_\gamma = 0.1$, we find that $\left|\rho_{tc}\right| \gtrsim 0.5$ ($\left|\rho_{tu}\right| \gtrsim 0.38$) is excluded at 95\% CL. The limit decreases for $c_\gamma < 0.1 $ and disappears for $c_\gamma = 0$, which we assume in our analysis. Note that $D$-$\bar D$ mixing and $B \to \tau \nu$ observables impose strong bounds on $\rho_{tu}$~\cite{Crivellin:2013wna}, requiring it to be very small; we set it to zero in our study. 

Before discussing LHC search for $A \to ZH$ in $\ell^+\ell^- tc$ final state at 14 TeV, we point out that the chosen benchmark model ($m_H = 200$ GeV, $|\rho_{tt}| = 0.3$, and $|\rho_{tc}| = 0.1$) evades all the constraints mentioned above.

\section{LHC search for $A \to ZH$}
We now discuss the LHC search prospects for the $A \to ZH \to \ell^+\ell^- t c$ channel, which leads to a final state consisting of three leptons plus at least two jets, with at least one $b$-tagged and one non-$b$-tagged, and missing transverse energy. 
We use the simple {\it cut and count} method and assume 14~TeV collision energy.
We choose $m_A = 450\,(500)$~GeV as a benchmark point, denoted BP\,(BP$^\prime$), where $\mathcal{B}(A \to ZH) \simeq 87\,(90)\%$.
This benchmark could also be used to search for charged Higgs bosons via $H^+ \to W^+ H$ \cite{Hou:2024ibt}.
Our signal can be searched for via $pp \to A \to ZH \to \ell^+\ell^- (t\bar c+ \bar t c)+ X$, with $t \to \ell^+\nu b$, where $\ell = e$ or $\mu$. 
The dominant SM backgrounds are $WZ$, $tZj$, $t\bar tZ$, $t\bar tW$, and $tZc$, with $WWZ$, $WZZ$, $t\bar th$, and $t\bar tt\bar t$ are subdominant. 
Other backgrounds such as $tttj$ and $tttW$ are minor. 

Signal and background samples are generated at leading order (LO) using \texttt{MadGraph5\_aMC@NLO}~\cite{Alwall:2014hca}.
These samples are passed to \texttt{Pythia-8.2}~\cite{Sjostrand:2014zea} for parton showering and hadronization, and then to \texttt{Delphes-3.5.0} \cite{deFavereau:2013fsa}, with a default ATLAS card for detector simulation.
We consider one additional jet for $WZ$, ${t\bar t}Z$, and ${t\bar t}W$ backgrounds using the MLM matching scheme~\cite{Alwall:2007fs}. The $tZj$, $tZc$, $WWZ$, $WZZ$, ${t\bar t}h$, $t\bar t t\bar t$ backgrounds and signal samples are estimated without additional jets. 

We rescale the LO cross sections for our signal and the dominant background $WZ$ to their respective next-to-next-to-leading-order values using $K$ factors of 2.14 (2.09) for BP (BP$^\prime$), and 2.07 (2.02) for $W^-Z$ ($W^+Z$)~\cite{Grazzini:2016swo}.
Note that the signal $K$-factor is computed using the code \texttt{SusHi}~\cite{Harlander:2012pb}.
The LO $tZj$, $t\bar t Z$, $t\bar t W^-$ ($t\bar t W^+$), $WWZ$, $WZZ$, $t\bar th$, and $t\bar tt\bar t$ backgrounds are adjusted to next-to-leading order by factors of 1.44~\cite{Alwall:2014hca}, 1.40~\cite{LHCHiggsCrossSectionWorkingGroup:2016ypw}, 1.54 (1.51)~\cite{LHCHiggsCrossSectionWorkingGroup:2016ypw}, 1.74~\cite{Alwall:2014hca}, 1.85~\cite{Alwall:2014hca}, 1.26~\cite{LHCHiggsCrossSectionWorkingGroup:2016ypw}, and 2.04~\cite{Alwall:2014hca}, respectively. The $tZc$ background is assumed to have the same correction factor as $tZj$. 

For event selection, we require the presence of exactly three charged leptons (samples with more than three leptons are vetoed), with $p^{\ell_1}_T > 80$, $p^{\ell_2}_T > 30$, and $p^{\ell_3}_T > 20$~GeV, and $|\eta_\ell|<2.5$, together with at least two jets, of which there is at least one $b$-tagged and one non-$b$-tagged, with $P^j_T > 20$~GeV and $|\eta_j|<2.5$, plus a missing transverse energy $E^{\rm{miss}}_{T}$ larger than 20~GeV. The scalar $p_T$ sum of all jets and the three charged leptons ($H_T$) is required to be between 280 and 500~GeV. To further suppress the $t\bar tW$, $t\bar th(\to WW^*)$, and $t\bar t t\bar t$ backgrounds, the invariant mass $m_{\ell^+\ell^-}$ of all possible same-flavor, opposite-sign $\ell^+ \ell^-$ pairs must lie within the mass window  $70 < m_{\ell^+\ell^-} < 110$~GeV.
Note that the event selection cuts, particularly those on $H_T$ and $P^{\ell}_T$, are chosen to maximize the signal significance.
We also note that reconstructing the top quark (from the $b$-tagged jet, the missing energy and the remaining lepton) could further suppress the dominant $WZ$ background, as well as the subdominant $WWZ$ and $WZZ$ backgrounds, which in turn would enhance the signal significance.

\begin{table}[b!]
	\centering
	\setlength{\tabcolsep}{15pt}
	{\small \begin{tabular}{l c} 
			\hline\hline
			Process & Cross section \\
			\hline
			BP\,(BP$^\prime$) & 0.87\,(0.53)\\
			$WZ$ & 0.81 \\			
			$tZj$ & 0.36 \\
			$t\bar tZ$ & 0.17 \\			
			$t\bar tW$ & 0.036 \\
			$tZc$ & 0.034 \\	
			$WWZ$ & 0.008 \\
			$WZZ$ & 0.007 \\							
			$t\bar th$ & 0.002 \\
			$t\bar t t\bar t$ & $< 0.001$ \\ 
			\hline\hline
	\end{tabular}}
	\caption{Signal and background cross sections (in fb) at 14~TeV after selection cuts.}
	\label{table:xsAC}
\end{table} 

The $P^{\ell_1}_T$, $H_T$, and $m_{\ell^+\ell^-}$ distributions, before any selection cuts (only default cuts during event generation in \texttt{MadGraph5\_aMC@NLO} are applied), are shown in Fig.~\ref{fig:distr}.
Signal and background cross sections, after selection cuts, are given in Table~\ref{table:xsAC}.
Note that the $t\bar t + \rm{jets}$ process can also contribute if a jet is misidentified as a lepton (fake). 
We adopt a conservative misidentification rate of $\epsilon_{\rm{fake}} = 10^{-4}$~\cite{ATLAS:2016dlg,Alvarez:2016nrz}
and find the resulting cross section is below 0.001 after selection cuts. 
Thus, we neglect the fake contribution in our analysis.

Based on the results given in Table~\ref{table:xsAC}, we determine the statistical significance for our signal with 
$\mathcal{Z} = \sqrt{2\left[(S+B)\ln(1+S/B)-S\right]}$~\cite{Cowan:2010js}, where $S$ ($B$) is the number of signal (background) events.
For 140 fb$^{-1}$, the significance for $A \to ZH \to \ell^+\ell^- t c$ is $\sim 7.9\sigma$ ($\sim 5.0\sigma$) for BP (BP$^\prime$). 
Thus, the available Run 2 data could lead to discovery. Assuming a conservative 10\% systematic uncertainty in the background estimation, we find 
a significance of $\sim 4.4\sigma$ ($\sim 2.8\sigma$) for BP (BP$^\prime$). 

\section{CONCLUSIONS}
ATLAS~\cite{ATLAS:2023tlp} and CMS~\cite{CMS:2023xpx} have recently conducted searches for G2HDM Higgs bosons, marking a remarkable step by exploring new channels beyond the conventional ones.
G2HDM is a quite robust model, as it can drive~\cite{Fuyuto:2017ewj} EWBG that could explain the observed matter-antimatter asymmetry of the Universe. 
It is found that a strong FOEWPT in 2HDM, as needed for EWBG, favors a rather heavy {\it CP}-odd Higgs boson with $m_A >$ 300 GeV and a large mass splitting $m_A - m_H \gtrsim v$ \cite{Dorsch:2014qja}. Hence, the decay $A \to ZH$ is identified as a smoking-gun signature of 2HDM scenarios with a strongly FOEWPT, and searches for $A \to ZH$ in $\ell^+\ell^- b\bar b$ and $\ell^+\ell^- W^+W^-$ final states are found promising~\cite{Dorsch:2014qja}. 
In addition, $A \to ZH$ can also be explored in the $\ell^+\ell^- t\bar t$ and $\nu\nu b\bar b$ final states~\cite{Biekotter:2023eil}.
In this paper, we have shown that $A \to ZH$ in the $\ell^+\ell^- t c$ final state is very promising, and the current Run 2 data may be able to discover this exotic decay through the trilepton signal. Observation of this final state would be a smoking-gun for the G2HDM.

We have assumed a scenario where $m_{A} \simeq m_{H^+}$ (custodial), but one can, on other hand, consider the $m_H \simeq m_{H^+}$ case (twisted custodial~\cite{Gerard:2007kn}), with $m_{H^+} < m_A$. For example, let us consider $m_H = m_{H^+} = 200$ GeV and $m_A = 500$ GeV (with $|\rho_{tt}| = 0.3,\ |\rho_{tc}| = 0.1$). In this benchmark scenario, the {\it CP}-odd Higgs boson $A$ would decay into $W^-H^+$, with a branching ratio of $\sim 65\%$, which dilutes the $A \to ZH$ decay, thereby altering the expected number of events in $\ell^+\ell^- t c$ final state. In addition, the decay $H^+ \to c\bar b$ would be enhanced, suggesting $A \to W^-H^+ \to \ell^-\bar \nu c\bar b$ could serve as an alternative discovery channel.
However, for Higgs masses above 200~GeV (e.g., $m_H \simeq m_{H^+} \simeq 300$ GeV), the $H^+ \to t\bar b$ decay would dominate, making the $A \to W^-H^+ \to \ell^-\bar \nu t\bar b$ final state potentially more promising.
It should be noted that our analysis can also be helpful to search for the $H \to ZA$ inverted scenario, which can also yield a FOEWPT (see, e.g., Refs.~\cite{Basler:2016obg, Bernon:2017jgv}; see also Ref.~\cite{Arco:2025ydq} for a recent study on distinguishing between the $A \to ZH$ and $H \to ZA$ signatures at the LHC).

Away from the alignment limit, $A \to Zh$, $H \to W^+W^-$, and $H \to ZZ$ would be induced. For $c_\gamma = 0.1$, which agrees with standard model Higgs measurements at $2\sigma$ level by \texttt{HiggsSignals}, the decay $A \to Zh$ remains negligible. However, $H \to W^+W^-$ ($ZZ$) is important and has a branching ratio of $\sim 35\%$ ($\sim 16\%$), diluting $H \to t\bar c + \bar tc$. 
For nonzero $c_\gamma$, $A \to ZH \to  \ell^+\ell^- t c$ and $A \to ZH \to  \ell^+\ell^- W^+W^-$ would both lead to a trilepton signal, which is itself another G2HDM bonus. Note that $A \to ZH \to  \ell^+\ell^- ZZ$ is also possible, but it would require much more data for potential discovery. 

So far, we have only considered extra top Yukawa couplings $\rho_{tt}$ and $\rho_{tc}$.
Turning on $\rho_{bb}~(\rho_{\tau\tau})$ would induce $H \to b\bar b~(\tau^+\tau^-)$, and thus one can have $A \to ZH \to \ell^+\ell^- b\bar b$. However, flavor observables strongly constrain $\rho_{bb}~(\rho_{\tau\tau})$, requiring $|\rho_{bb}| \lesssim \lambda_{bb}~(|\rho_{\tau\tau}| \lesssim \lambda_{\tau\tau})$~\cite{Crivellin:2013wna,Altunkaynak:2015twa}. For $\rho_{bb} \sim \rho_{\tau\tau} \sim 0.01$, $H \to b\bar b$ and $H \to \tau^+\tau^-$ are small, and their presence would slightly soften $H \to t\bar c + \bar tc$ but would not alter our conclusions. Note that nonzero $\rho_{bb}$ and $\rho_{\tau\tau}$ would also induce $A \to b\bar b$ and $A \to \tau^+\tau^-$, but these decays are very small.

Other flavor-changing couplings, such as $\rho_{\mu\tau}$ and $\rho_{\tau\mu}$, come to mind; however, these couplings are severely constrained by lepton-flavor-violating $\tau^- \to \mu^- \mu^+\mu^-$ decay. For $m_{H} = 500$ GeV $(\approx m_{H^+} \approx m_A)$, $|\rho_{\mu\tau, \tau\mu}| \leq 3.7 \times 10^{-3}$~\cite{Crivellin:2013wna}.
Hence, $\rho_{\mu\tau}$ and $\rho_{\tau\mu}$ must be very small.
Note that this is only an example; other off-diagonal couplings are also strongly constrained.
The $\rho_{\mu\tau}$ and $\rho_{\tau\mu}$ couplings are further constrained by flavor-violating Higgs boson $h \to \mu\tau$ decay searches~\cite{CMS:2021rsq}. 
These bounds disappear in the limit of $c_\gamma \to 0$ (alignment), which we assume to ensure agreement with Higgs signal measurements. 
Of course, nonzero $\rho_{\mu\tau}$ and/or $\rho_{\tau\mu}$ would induce $H/A \to \mu\tau$ but these decays are very small. The $H \to t\bar c + \bar t c$ and $A \to ZH$ decays can still be the dominant channels, and therefore our results remain unchanged. 
For the same reason, these flavor-changing couplings and even $\rho_{e\mu}$, $\rho_{\mu e}$, $\rho_{bs}$, or $\rho_{bd}$ are turned off. 

In summary, we suggest searching for $A \to ZH$ in the $\ell^+\ell^- t c$ final state, a smoking-gun signature for G2HDM. We proposed a benchmark scenario and carried out a signal-to-background analysis at the 14 TeV LHC and show that $A \to ZH$ is quite promising. Even current LHC Run 2 data may give potential discovery of this exotic decay into $\ell^+\ell^- t c$ final state, thus providing a probe of G2HDM with flavor violation.

\vskip0.2cm
\noindent{\bf Acknowledgments.--}
This work is supported by NSTC Grant No. 113-2639-M-002-006-ASP of Taiwan, and NTU Grants No. 113L86001 and No. 113L891801. M.~K. acknowledges CERN TH Department for hospitality, where part of this work was carried out.



\begin{thebibliography}{99}
\bibitem{ATLAS:2012yve}
G.~Aad \textit{et al.} [ATLAS],
Phys. Lett. B \textbf{716}, 1 (2012).

\bibitem{CMS:2012qbp}
S.~Chatrchyan \textit{et al.} [CMS],
Phys. Lett. B \textbf{716}, 30 (2012).

\bibitem{Kanemura:2004ch}
S.~Kanemura, Y.~Okada, and E.~Senaha,
Phys. Lett. B \textbf{606}, 361 (2005).

\bibitem{Fuyuto:2017ewj}
K.~Fuyuto, W.-S.~Hou and E.~Senaha,
Phys. Lett. B \textbf{776}, 402 (2018).


\bibitem{Turok:1990zg}
N.~Turok and J.~Zadrozny,
Nucl. Phys. B \textbf{358}, 471 (1991).

\bibitem{Davies:1994id}
A.~T.~Davies, C.~D.~froggatt, G.~Jenkins and R.~G.~Moorhouse,
Phys. Lett. B \textbf{336}, 464 (1994).

\bibitem{Cline:1995dg}
J.~M.~Cline, K.~Kainulainen and A.~P.~Vischer,
Phys. Rev. D \textbf{54}, 2451 (1996).

\bibitem{Cline:1996mga}
J.~M.~Cline and P.~A.~Lemieux,
Phys. Rev. D \textbf{55}, 3873 (1997).

\bibitem{Fromme:2006cm}
L.~Fromme, S.~J.~Huber and M.~Seniuch,
JHEP \textbf{11} (2006) 038.

\bibitem{Cline:2011mm}
J.~M.~Cline, K.~Kainulainen and M.~Trott,
JHEP \textbf{11} (2011) 089.

\bibitem{Ahmadvand:2013sna}
M.~Ahmadvand,
Int. J. Mod. Phys. A \textbf{29}, 1450090 (2014).

\bibitem{Dorsch:2013wja}
G.~C.~Dorsch, S.~J.~Huber and J.~M.~No,
JHEP \textbf{10} (2013) 029.

\bibitem{Dorsch:2014qja}
G.~C.~Dorsch, S.~J.~Huber, K.~Mimasu and J.~M.~No,
Phys. Rev. Lett. \textbf{113}, 211802 (2014).

\bibitem{Basler:2016obg}
P.~Basler, M.~Krause, M.~Muhlleitner, J.~Wittbrodt and A.~Wlotzka,
JHEP \textbf{02} (2017) 121.

\bibitem{Bernon:2017jgv}
J.~Bernon, L.~Bian and Y.~Jiang,
JHEP \textbf{05} (2018) 151.

\bibitem{Biekotter:2023eil}
T.~Biek\"otter, S.~Heinemeyer, J.~M.~No, K.~Radchenko, M.~O.~O.~Romacho and G.~Weiglein,
JHEP \textbf{01} (2024) 107.

\bibitem{Chiang:2016vgf}
C.~W.~Chiang, K.~Fuyuto and E.~Senaha,
Phys. Lett. B \textbf{762}, 315 (2016).

\bibitem{Dorsch:2016nrg}
G.~C.~Dorsch, S.~J.~Huber, T.~Konstandin and J.~M.~No,
JCAP \textbf{05} (2017) 052.

\bibitem{Dorsch:2017nza}
G.~C.~Dorsch, S.~J.~Huber, K.~Mimasu and J.~M.~No,
JHEP \textbf{12} (2017) 086.

\bibitem{Modak:2018csw}
T.~Modak and E.~Senaha,
Phys. Rev. D \textbf{99}, 115022 (2019).

\bibitem{Goncalves:2021egx}
D.~Gon\c{c}alves, A.~Kaladharan and Y.~Wu,
Phys. Rev. D \textbf{105}, 095041 (2022).

\bibitem{Basler:2021kgq}
P.~Basler, L.~Biermann, M.~M\"uhlleitner and J.~M\"uller,
Eur. Phys. J. C \textbf{83}, 57 (2023).

\bibitem{Enomoto:2021dkl}
K.~Enomoto, S.~Kanemura and Y.~Mura,
JHEP \textbf{01} (2022) 104;
K.~Enomoto, S.~Kanemura and Y.~Mura,
JHEP \textbf{09} (2022) 121.

\bibitem{Kanemura:2023juv}
S.~Kanemura and Y.~Mura,
JHEP \textbf{09} (2023) 153.

\bibitem{Goncalves:2023svb}
D.~Gon\c{c}alves, A.~Kaladharan and Y.~Wu,
Phys. Rev. D \textbf{108}, 075010 (2023).

\bibitem{Athron:2025iew}
P.~Athron, M.~J.~Ramsey-Musolf, C.~Sierra and Y.~Wu,
[arXiv:2502.00445 [hep-ph]].

\bibitem{ACME:2018yjb}
V.~Andreev \textit{et al.} (ACME Collaboration),
Nature \textbf{562}, 355 (2018).

\bibitem{Roussy:2022cmp}
T.S.~Roussy, \textit{et al.}
Science \textbf{381}, 46 (2023).

\bibitem{Fuyuto:2019svr}
K.~Fuyuto, W.-S.~Hou and E.~Senaha,
Phys. Rev. D \textbf{101}, 011901(R) (2020).

\bibitem{Hou:1991un}
W.-S.~Hou,
Phys. Lett. B \textbf{296}, 179 (1992).

\bibitem{Kohda:2017fkn}
M.~Kohda, T.~Modak, and W.-S.~Hou,
Phys. Lett. B \textbf{776}, 379 (2018).

\bibitem{Ghosh:2019exx}
D.K.~Ghosh, W.-S.~Hou, and T.~Modak,
Phys. Rev. Lett. \textbf{125}, 221801 (2020).

\bibitem{Hou:2024bzh}
W.-S.~Hou and M.~Krab,
Phys. Rev. D \textbf{110}, L011702 (2024).

\bibitem{Hou:2024ibt}
W.-S.~Hou and M.~Krab,
Phys. Rev. D \textbf{111}, L031701 (2025).

\bibitem{Davidson:2005cw}
S.~Davidson and H.E.~Haber,
Phys.\ Rev.\ D {\bf 72}, 035004 (2005).

\bibitem{Hou:2017hiw}
W.-S.~Hou and M.~Kikuchi,
EPL \textbf{123}, 11001 (2018).

\bibitem{ATLAS:2016neq}
G.~Aad \textit{et al.} [ATLAS and CMS],
JHEP \textbf{08} (2016) 045.

\bibitem{Glashow:1976nt}
S.~L.~Glashow and S.~Weinberg,
Phys. Rev. D \textbf{15}, 1958 (1977).

\bibitem{Gerard:2007kn}
{J.-M.~G\'erard} and M.~Herquet,
Phys. Rev. Lett. \textbf{98}, 251802 (2007).

\bibitem{Eriksson:2009ws}
D.~Eriksson, J.~Rathsman and O.~Stål,
Comput. Phys. Commun. \textbf{181}, 189 (2010).

\bibitem{Grimus:2007if}
W.~Grimus, L.~Lavoura, O.~M.~Ogreid and P.~Osland,
J. Phys. G \textbf{35}, 075001 (2008);
Nucl. Phys. B \textbf{801}, 81 (2008).

\bibitem{ParticleDataGroup:2024cfk}
S.~Navas \textit{et al.} [Particle Data Group],
Phys. Rev. D \textbf{110}, 030001 (2024).

\bibitem{Crivellin:2013wna}
A.~Crivellin, C.~Greub and A.~Kokulu,
Phys. Rev. D \textbf{87}, 094031 (2013).

\bibitem{Altunkaynak:2015twa}
B.~Altunkaynak, W.-S.~Hou, C.~Kao, M.~Kohda and B.~McCoy,
Phys. Lett. B \textbf{751}, 135 (2015).

%
\bibitem{UTFit2023}
\url{http://www.utfit.org/UTfit/ResultsSummer2023NP}.
%

\bibitem{Hou:2018uvr}
W.-S.~Hou, M.~Kohda and T.~Modak,
Phys. Rev. D \textbf{98}, 075007 (2018).

\bibitem{Bechtle:2020uwn}
P.~Bechtle, S.~Heinemeyer, T.~Klingl, T.~Stefaniak, G.~Weiglein and J.~Wittbrodt,
Eur. Phys. J. C \textbf{81}, 145 (2021).

\bibitem{Bahl:2022igd}
H.~Bahl, T.~Biek\"otter, S.~Heinemeyer, C.~Li, S.~Paasch, G.~Weiglein and J.~Wittbrodt,
Comput. Phys. Commun. \textbf{291}, 108803 (2023).

\bibitem{ATLAS:2021upq}
G.~Aad \textit{et al.} [ATLAS],
JHEP \textbf{06} (2021) 145.

\bibitem{CMS:2020imj}
A.M.~Sirunyan \textit{et al.} [CMS],
JHEP \textbf{07} (2020) 126.

\bibitem{ATLAS:2023tlp}
G.~Aad \textit{et al.} [ATLAS],
JHEP \textbf{12} (2023) 081.

\bibitem{CMS:2023xpx}
A.~Hayrapetyan \textit{et al.} [CMS],
Phys. Lett. B \textbf{850}, 138478 (2024).

\bibitem{CMS:2021hug}
A.~Tumasyan \textit{et al.} [CMS],
Phys. Rev. Lett. \textbf{129}, 032001 (2022);
A.~Hayrapetyan \textit{et al.} [CMS],
arXiv:2407.15172.

\bibitem{ATLAS:2024mih}
G.~Aad \textit{et al.} [ATLAS],
Eur. Phys. J. C \textbf{84}, 757 (2024).

\bibitem{Alwall:2014hca}
J.~Alwall {\it et al.}, 
JHEP \textbf{07} (2014) 079.

\bibitem{Sjostrand:2014zea}
T.~Sj\"ostrand {\it et al.}, 
Comput. Phys. Commun. \textbf{191}, 159 (2015).

\bibitem{deFavereau:2013fsa}
J.~de Favereau \textit{et al.} [DELPHES 3],
JHEP \textbf{02} (2014) 057.

\bibitem{Alwall:2007fs}
J.~Alwall \textit{et al.}
Eur. Phys. J. C \textbf{53}, 473 (2008).

\bibitem{Grazzini:2016swo}
M.~Grazzini, S.~Kallweit, D.~Rathlev and M.~Wiesemann,
Phys. Lett. B \textbf{761}, 179 (2016).

\bibitem{Harlander:2012pb}
R.~V.~Harlander, S.~Liebler and H.~Mantler,
Comput. Phys. Commun. \textbf{184}, 1605 (2013).

\bibitem{LHCHiggsCrossSectionWorkingGroup:2016ypw}
D.~de Florian \textit{et al.} [LHC Higgs Cross Section Working Group],
arXiv:1610.07922.

\bibitem{ATLAS:2016dlg}
G.~Aad \textit{et al.} [ATLAS],
Eur. Phys. J. C \textbf{76}, 259 (2016).

\bibitem{Alvarez:2016nrz}
E.~Alvarez, D.A.~Faroughy, J.F.~Kamenik, R.~Morales, and A.~Szynkman,
Nucl. Phys. B \textbf{915}, 19 (2017).

\bibitem{Cowan:2010js}
G.~Cowan, K.~Cranmer, E.~Gross and O.~Vitells,
Eur. Phys. J. C \textbf{71}, 1554 (2011).

\bibitem{Arco:2025ydq}
F.~Arco, T.~Biek\"otter, P.~Stylianou and G.~Weiglein,
[arXiv:2502.03443 [hep-ph]].

\bibitem{CMS:2021rsq}
A.~M.~Sirunyan \textit{et al.} [CMS],
Phys. Rev. D \textbf{104}, 032013 (2021).

\end{thebibliography}
\end{document}